\documentclass[aps,prl,twocolumn,preprintnumbers,nobalancelastpage,showpacs]{revtex4}

\begin{document}

\title{Extreme Fluctuations in Small-Worlds with Relaxational Dynamics}

\author{H. Guclu}
\email{gucluh@rpi.edu}    
\author{G. Korniss}
\email{korniss@rpi.edu}

\affiliation{Department of Physics, Applied
Physics, and Astronomy, Rensselaer Polytechnic Institute, 110
8$^{th}$ Street, Troy, NY 12180--3590, USA}

%%%%%%%%%%%%%%%%%%%%%%%%%%%%%%%%%%%%%%%%%%%%%%%%%%%%%%%%%%%%%%%%%%%

\begin{abstract}
{We study the distribution and scaling of the extreme height
fluctuations for Edwards-Wilkinson-type relaxation on small-world substrates. 
When random links are added to a one-dimensional lattice, the average
size of the fluctuations becomes finite (synchronized state) and the
extreme height diverges only logarithmically in the large system-size limit. 
This latter property ensures synchronization in a
practical sense in small-world coupled multi-component autonomous
systems. The statistics of the extreme heights is governed by the
Fisher-Tippett-Gumbel distribution.}
\end{abstract}


\date{\today}
\maketitle

Synchronization is a fundamental problem in natural and artificial coupled
multi-component systems \cite{Strogatz_review}. Since the introduction
of small-world (SW) networks \cite{WATTS98} it has been well established that
such networks can facilitate autonomous synchronization
\cite{Strogatz_review}. Examples include noisy coupled phase
oscillators \cite{phase_sw} and scalable parallel simulators for
asynchronous dynamics \cite{KORNISS03a}. In
essence, the SW coupling introduces an effective
relaxation to the mean of the respective local field variables (or local
``load''), and induces (strict or anomalous) mean-field-like behavior
\cite{HASTINGS03,KOZMA03}. In addition to the average load in the
network, knowing the typical size and the distribution of the extreme  
fluctuations \cite{FT,GUMBEL,GALAMBOS} is of great importance from a
system-design viewpoint, since failures and delays are triggered by
extreme events occurring on an individual node.

In this Letter, we focus on the steady-state properties of the extreme
fluctuations in SW-coupled interacting systems with relaxational dynamics.
In contrast, consider, for example,
kinetically growing possibly non-equilibrium surfaces with only
short-range interactions (e.g., nearest neighbors on a lattice).
Here a suitably chosen local field variable is the local height
fluctuation measured from the mean \cite{BARABASI}. It was shown
\cite{SHAPIR} that in the steady state, where the surface is rough,
the extreme height fluctuations diverge in the same power-law 
fashion with the system size as the average height fluctuations
(the width). Similar observation was made
\cite{KORNISS_ACM} in the context of the scalability
of parallel discrete-event simulations (PDES)
\cite{FUJI,KORNISS00,LUBA}, where the
progress of the simulation is governed by the Kardar-Parisi-Zhang (KPZ)
equation \cite{KPZ}: here the ``relative height'' or local field
variable is the deviation of the progress of the individual
processor from the average rate of progress of the simulation
\cite{KORNISS00}. 
The systems in the above examples are ``critical'' in
that the lateral correlation length of the corresponding rough  
surfaces scales with the system size \cite{BARABASI}. For systems
at criticality with unbounded local variables, the extreme
values of the local fluctuations emerge through the dominating
collective long-wavelength modes, and the extremal and the average fluctuations
follow the same power-law divergence with the system size
\cite{SHAPIR,KORNISS_ACM}. Relationship between extremal
statistics and universal fluctuations in correlated systems
have been studied intensively
\cite{SHAPIR,BRAMWELL_NATURE,BRAMWELL_PRL,GOLD,ADGR,CHAPMAN,BOUCHAUD}.
Here we discuss to what extent SW couplings
(extending the original dynamics through the random links)
lead to the suppression of the extreme fluctuations. We illustrate our
findings on an actual synchronization problem for scalable PDES
schemes \cite{KORNISS03a}.

First, we briefly summarize the basic properties of the extremal values of
$N$ independent stochastic variables \cite{FT,GUMBEL,GALAMBOS,AB,BOUCHAUD}.
Here we consider the case when the individual complementer cumulative
distribution $P_{>}(x)$ (the probability that the individual
stochastic variable is greater than $x$) decays faster than any power
law, i.e., exhibits an exponential-like tail in the asymptotic
large-$x$ limit. (Note that in this case the corresponding probability density
function displays the same exponential-like asymptotic tail behavior.)  
We will assume $P_{>}(x)\simeq e^{-cx^{\delta}}$ 
for large $x$ values, where $c$ is a constant. 
Then the cumulative distribution $P_{<}^{\max}(x)$
for the largest of the $N$ events (the probability that the maximum
value is less than $x$) can be approximated as \cite{BOUCHAUD,AB}
%%%%%%%%%%%%%%%%%%%%%%%%%%%%%%%%%%%%%%%%%%%%%%%%%%%%%%%%%%%%%%%%%%%%%%%
\begin{equation}
P_{<}^{\max}(x) = [P_{<}(x)]^N = [1-P_{>}(x)]^N \simeq e^{-NP_{>}(x)} \;,
\end{equation}
%%%%%%%%%%%%%%%%%%%%%%%%%%%%%%%%%%%%%%%%%%%%%%%%%%%%%%%%%%%%%%%%%%%%%%%
where one typically assumes that the dominant contribution to the
statistics of the extremes comes from the tail of the individual
distribution $P_{>}(x)$. With the exponential-like tail in the
individual distribution, this yields
%%%%%%%%%%%%%%%%%%%%%%%%%%%%%%%%%%%%%%%%%%%%%%%%%%%%%%%%%%%%%%%%%%%%%%%
\begin{equation}
P_{<}^{\max}(x) \simeq e^{-e^{-cx^{\delta}+\ln(N)}}.
\label{raw_extreme}
\end{equation}
%%%%%%%%%%%%%%%%%%%%%%%%%%%%%%%%%%%%%%%%%%%%%%%%%%%%%%%%%%%%%%%%%%%%%%%
The extreme-value limit theorem states that there exists a sequence of
scaled variables $\tilde{x}=(x-a_{N})/b_{N}$, such that in the limit of
$N$$\to$$\infty$, the extreme-value probability distribution for $\tilde{x}$
asymptotically approaches the Fisher-Tippett-Gumbel (FTG)
distribution \cite{FT,GUMBEL}:
%%%%%%%%%%%%%%%%%%%%%%%%%%%%%%%%%%%%%%%%%%%%%%%%%%%%%%%%%%%%%%%%%%%%%%%
\begin{equation}
\tilde{P}_{<}^{\max}(\tilde{x}) \simeq e^{-e^{-\tilde{x}}} \;,
\label{gumbel}
\end{equation}
%%%%%%%%%%%%%%%%%%%%%%%%%%%%%%%%%%%%%%%%%%%%%%%%%%%%%%%%%%%%%%%%%%%%%%%
with mean $\langle\tilde{x}\rangle$$=$$\gamma$ 
($\gamma$$=$$0.577\ldots$ being the Euler constant) and variance
$\sigma_{\tilde{x}}^{2}$$=$$\langle\tilde{x}^{2}\rangle$$-$$\langle\tilde{x}\rangle^{2}$$=$$\pi^2/6$. 
From Eq.~(\ref{raw_extreme}), one can deduce that to leading
order the scaling coefficients must be $a_{N}$$=$$[\ln(N)/c]^{1/\delta}$ and
$b_{N}$$=$$(\delta c)^{-1}[\ln(N)/c]^{(1/\delta)-1}$ \cite{AB,note1}.
The average value of the largest of the $N$ original variables then scales as
%%%%%%%%%%%%%%%%%%%%%%%%%%%%%%%%%%%%%%%%%%%%%%%%%%%%%%%%%%%%%%%%%%%%%%%
\begin{equation}
\langle x_{\max}\rangle = a_N + b_N\gamma \simeq [\ln(N)/c]^{1/\delta}
\label{mean}
\end{equation}
%%%%%%%%%%%%%%%%%%%%%%%%%%%%%%%%%%%%%%%%%%%%%%%%%%%%%%%%%%%%%%%%%%%%%%%
(up to ${\cal O}(\frac{1}{\ln(N)})$ correction) in the asymptotic
large-$N$ limit.

We now consider the Edwards-Wikinson (EW) model \cite{EW},
a prototypical synchronization problem with
relaxational dynamics \cite{KORNISS03a,KORNISS00,KOZMA03},
first on a regular one-dimensional lattice:
%%%%%%%%%%%%%%%%%%%%%%%%%%%%%%%%%%%%%%%%%%%%%%%%%%%%%%%%%%%%%%%%%%%%%%%%%
$\partial_t h_i=-\sum_{j}\Gamma^o_{ij} h_j + \eta_i(t)$.
%%%%%%%%%%%%%%%%%%%%%%%%%%%%%%%%%%%%%%%%%%%%%%%%%%%%%%%%%%%%%%%%%%%%%%%%%
Here $h_i$ is the local height (or load) at site $i$,
$-\Gamma^o_{ij}$$=$$\delta_{i j\!+\!1}+\delta_{i j\!-\!1}-2\delta_{ij}$
is the discrete Laplacian on the lattice, and $\eta_i(t)$ is a
short-tailed noise (e.g., Gaussian), delta-correlated in space and time. 
The width, borrowing the framework from non-equilibrium surface-growth
phenomena, provides a sensitive measure for the average degree of
synchronization in coupled multi-component systems
\cite{KORNISS03a,KORNISS00}.
The EW model on a one-dimensional lattice with $N$ sites has a
``rough'' surface profile in the steady state (de-synchronized state),
where the average width
%%%%%%%%%%%%%%%%%%%%%%%%%%%%%%%%%%%%%%%%%%%%%%%%%%%%%%%%%%%%%%%%%%%%%
$\langle w^2 \rangle_{N}$$\equiv$$%
\langle\frac{1}{N}\sum_{i=1}^{N}(h_i-\bar{h})^2\rangle$
%%%%%%%%%%%%%%%%%%%%%%%%%%%%%%%%%%%%%%%%%%%%%%%%%%%%%%%%%%%%%%%%%%%%%
diverges in the thermodynamic limit as $\langle w^2\rangle$$\sim$$N$.
(Here, $\bar{h}$$=$$(1/N)\sum_{i=1}^{N}h_i$ is the mean height.) 
The diverging width is related to an underlying diverging lengthscale, 
the lateral correlation length, which reaches the system size $N$ for a
finite system.
Further, the maximum height fluctuations (measured from the mean
height $\bar{h}$) diverges in the same fashion as the width itself,
i.e., $\langle (h_{\max}- \bar{h})^2\rangle$$\sim$$N$ 
\cite{SHAPIR,KORNISS_ACM}.

Here we ask how the scaling behavior of the extremal height
fluctuations changes if we {\em extend the same dynamics} to a
SW network \cite{WATTS98}.  Then the equation of motion becomes
%%%%%%%%%%%%%%%%%%%%%%%%%%%%%%%%%%%%%%%%%%%%%%%%%%%%%%%%%%%%%%%%%%%%%%%%%
$\partial_t h_i=-\sum_{j}(\Gamma^o_{ij}+V_{ij}) h_j + \eta_i(t)$,
%%%%%%%%%%%%%%%%%%%%%%%%%%%%%%%%%%%%%%%%%%%%%%%%%%%%%%%%%%%%%%%%%%%%%%%%%
where $-V_{ij}=J_{ij}-\delta_{ij}\sum_{l}J_{il}$ is the
Laplacian on the random part of the network. The symmetric matrix
$J_{ij}$ represents the {\em quenched} random links on top of the regular
lattice, i.e., it is $1$ if site $i$ and $j$ are connected and $0$ otherwise.
In a frequently studied
version of the SW network \cite{NEWMAN,NEWMAN_WATTS,MONA}
random links of unit strength are added to
all possible pair of sites with probability $p/N$ (``soft''
network \cite{KOZMA03}). Here $p$ becomes the average number of random
links per site. In a somewhat different
construction of the network (``hard'' network \cite{KORNISS03a,KOZMA03}), each
site has exactly one random link (i.e., pairs of sites are selected at
random, and once they are linked, they cannot be selected again) and
the strength of the 
interaction through the random link is $p$. 
Note that both SW constructions have a finite average degree for
each node, and are embedded in a finite dimension.
The common feature in both SW versions is an effective nonzero
mass $\Sigma(p)$ (in a field theory sense), generated by the
quenched-random structure \cite{KOZMA03}. In turn,
both the correlation length $\xi$$\simeq$$[\Sigma(p)]^{-1/2}$ and the
width $\langle w^2 \rangle$$\simeq$$(1/2)[\Sigma(p)]^{-1/2}$ approach
a finite value (synchronized state) and become self-averaging in the
$N$$\to$$\infty$ limit.
For example, for the EW model on the soft and hard SW versions in
one dimension, for small $p$ values, $\Sigma(p)$$\sim$$p^2$ and
$\Sigma(p)$$\sim$$p$, respectively \cite{KOZMA03}. 
Thus, the correlation length
becomes {\em finite} for an arbitrarily small but nonzero density of
random links (soft network), or for an arbitrarily small but nonzero
strength of the random links (one such link per site, hard network).

This is the fundamental effect of extending the original dynamics to a SW
network: it decouples the fluctuations of the originally correlated
system. Then, the extreme-value limit theorems can be applied using
the number of independent blocks $N/\xi$ in the system \cite{BOUCHAUD,AB}. 
Further, if the tail
of the noise distribution decays in an exponential-like fashion, the
individual relative height distribution will also do so \cite{note3}, and
depends on the combination
$\Delta_i/w$, where $\Delta_i$$=$$h_i$$-$$\bar{h}$ is the relative height
measured from the mean at site $i$, and 
$w$$\equiv$$\sqrt{\langle w^2\rangle}$. Considering, e.g., the
fluctuations above the mean for the individual sites, we will then have 
$P_{>}(\Delta_i)\simeq\exp[-c(\Delta_i/w)^{\delta}]$,
where $P_{>}(\Delta_i)$ denotes the ``disorder-averaged'' 
(averaged over network realizations) single-site relative height
distribution, which becomes independent of the site $i$ for SW networks. 
From the above it follows that the cumulative distribution for the 
extreme-height fluctuations relative to the mean 
$\Delta_{\max}$$=$$h_{\max}$$-$$\bar{h}$,
if scaled appropriately, will be given by Eq.~(\ref{gumbel}) \cite{note2}
in the asymptotic large-$N$ limit (such that $N/\xi$$\gg$$1$). Further, from 
Eq.~(\ref{mean}), the average maximum relative height will scale as 
%%%%%%%%%%%%%%%%%%%%%%%%%%%%%%%%%%%%%%%%%%%%%%%%%%%%%%%%%%%%%%%%%%%%%%%
\begin{equation}
\langle \Delta_{\max}\rangle \simeq
w \left(\frac{\ln(N/\xi)}{c}\right)^{1/\delta} 
\simeq
\frac{w}{c^{1/\delta}}\left[\ln(N)\right]^{1/\delta} \;,
\label{mean_max}
\end{equation}
%%%%%%%%%%%%%%%%%%%%%%%%%%%%%%%%%%%%%%%%%%%%%%%%%%%%%%%%%%%%%%%%%%%%%%%
where we dropped all $N$-independent terms. Note, that both $w$ and
$\xi$ approach their {\em finite} asymptotic $N$-independent values for
SW-coupled systems, and the only $N$-dependent factor is $\ln(N)$ for
large $N$ values. Also, the same logarithmic scaling with $N$ holds for
the largest relative deviations below the mean 
$\langle\bar{h}$$-$$h_{\min}\rangle$ 
and for the maximum spread $\langle h_{\max}$$-$$h_{\min}\rangle$.
This is the central point of our Letter: in small-world synchronized
systems with unbounded local variables driven by exponential-like noise
distribution (such as Gaussian), the extremal fluctuations increase only {\em
logarithmically} with the number of nodes. 
This weak divergence, which one can regard as
marginal, ensures synchronization
for practical purposes in coupled multi-component systems.

Next we illustrate our arguments on a SW-synchronized system, the scalable
parallel discrete-event simulator for asynchronous dynamics \cite{KORNISS03a}. 
Consider an arbitrary one-dimensional system with nearest-neighbor
interactions, in which the discrete events (update attempts in the
local configuration) exhibit Poisson asynchrony. In the
one site per processing element (PE) scenario, each PE has its
own local simulated time, constituting the virtual time horizon
$\{h_{i}(t)\}_{i=1}^{N}$ (essentially the progress of the individual nodes).
Here $t$ is the discrete number of
parallel steps executed by all PEs, which is proportional to the
wall-clock time and $N$ is the number of PEs. According to the
basic conservative synchronization scheme \cite{LUBA}, at each
parallel step $t$, only those PEs for which the local simulated
time is not greater then the local simulated times of their
virtual neighbors, can increment their local time by an
exponentially distributed random amount. (Without loss of
generality we assume that the mean of the local time increment is
one in simulated time units [stu].) Thus, denoting the virtual
neighborhood of PE $i$ by $S_{i}$, if $h_{i}(t)\leq\min_{j\epsilon
S_{i}}\{h_{j}(t)\}$, PE $i$ can update the configuration of the
underlying site it carries and determine the time of the next
event. Otherwise, it idles. Despite its simplicity, this rule
preserves unaltered the asynchronous causal dynamics of the
underlying system \cite{LUBA,KORNISS00}. In the original algorithm \cite{LUBA},
the virtual communication topology between PEs mimics the interaction
topology of the underlying system. For example, for a one-dimensional
system with 
nearest-neighbor interactions, the virtual neighborhood of PE $i$,
$S_{i}$, consists of the left and right neighbor, PE $i$$-$$1$ and
PE $i$$+$$1$. It was shown \cite{KORNISS00} that then the virtual
time horizon exhibits KPZ-like kinetic roughening and the
steady-state behavior in one dimension is governed by the EW
Hamiltonian. Thus, the average width of the virtual time
horizon (the spread in the progress of the individual PEs) diverges as
$N$$\to$$\infty$ \cite{KORNISS00},
hindering efficient data collection in the measurement phase of
the simulation \cite{KORNISS_ACM}. To achieve a near-uniform progress
of the PEs
{\em without} employing frequent global synchronizations, it was shown
\cite{KORNISS03a} that including randomly chosen PEs ({\em in
addition} to the nearest neighbors) in the virtual neighborhood
results in a finite average width. Here we demonstrate that SW
synchronization in the above PDES scheme results in 
logarithmically increasing extreme fluctuations in the simulated time
horizon, governed by the FTG distribution. 

In one implementation of the above scheme, each PE has exactly one
random neighbor (in addition to the nearest neighbors) and the local
simulated time of the random neighbor is checked only with probability
$p$ at every simulation step. The 
corresponding communication network is the hard version of the SW
network, as described earlier, and the ``strength'' of the random links is
controlled by the relative frequency $p$ of the basic synchronizational
steps through those links.
In an alternative implementation, the communication topology is the
soft version the SW network (with $p$ average number of random links
per PE), and the random neighbors are checked at every simulation
step together with the nearest neighbors. Note that in both
implementations (where the virtual neighborhood $S_{i}$ now includes
possible random neighbor(s) for each site $i$), the extra checking of
the simulated time of the random neighbor is not needed for the
faithfulness of the simulation. It is merely introduced to control the
width of the time horizon \cite{KORNISS03a}. 
%%%%%%%%%%%%%%%%%%%%%%%%%%%%%%%%%%%%%%%%%%%%%%%%%%%%%%%%%%%%%%%%%%%%%
\begin{figure}[t]
\vspace*{4.8truecm}
       \includegraphics{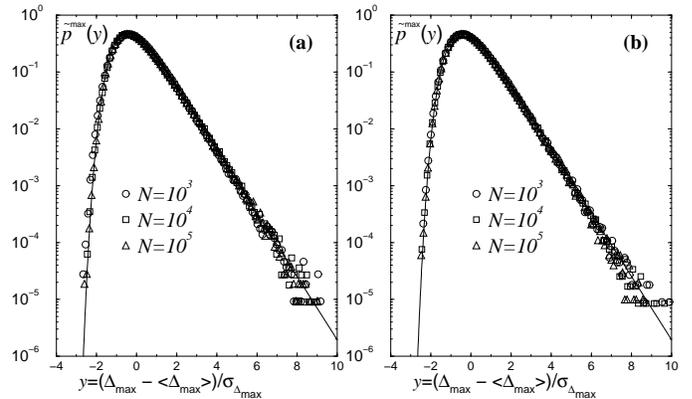}
\vspace*{0.5truecm}
\caption{Disorder-averaged probability densities for the scaled (zero
mean and unit variance) 
extreme-height fluctuations for the SW-synchronized PDES
time horizons. 
(a) For the hard SW network with $p$$=$$0.10$ and 
(b) for the soft SW network with $p$$=$$0.50$, on log-normal scales for three
system sizes. The solid curve corresponds to the similarly scaled FTG
density \cite{note2} for comparison.}
\label{fig1}
\end{figure}
%%%%%%%%%%%%%%%%%%%%%%%%%%%%%%%%%%%%%%%%%%%%%%%%%%%%%%%%%%%%%%%%%%%%

\begin{figure}[t]
\vspace*{2.1truecm}
       \includegraphics{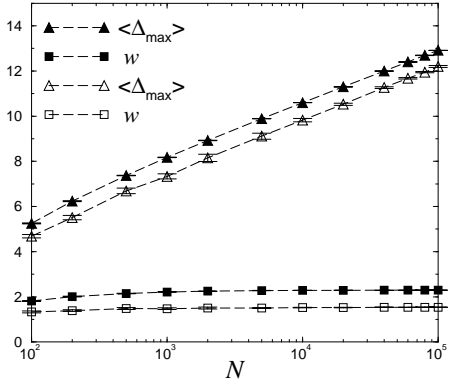}
\vspace*{3.0truecm}
\caption{Average maximum relative height and average width for the
SW-synchronized time horizon as a function of the number of nodes
(PEs). Filled symbols: hard SW network with $p$$=$$0.10$. Open symbols:
soft SW network with $p$$=$$0.50$. Note the normal-log scales.}
\label{fig2}
\end{figure}

To study the extreme fluctuations of the SW-synchronized virtual
time-horizon, we ``simulated the simulations'', i.e., the evolution of
the local simulated times based on the above exact algorithmic rules.
By constructing histograms for $\Delta_i$, we observed that the
tail of the disorder-averaged individual relative-height distribution decays
exponentially for both SW constructions. Then, we constructed histograms
for the scaled extreme-height fluctuations. The results, together with
the similarly scaled FTG density \cite{note2}, are shown in Fig.~\ref{fig1}.
We also observed that the distribution of the extreme values becomes
{\em self-averaging}, i.e., independent of the network realization.
Finally, Fig.~\ref{fig2} shows that for sufficiently
large $N$ (such that $w$ essentially becomes system-size independent)
the average (or typical) size of the extreme-height fluctuations
diverge {\em logarithmically}, according to Eq.~(\ref{mean_max}) with
$\delta$$=$$1$. We also found that the largest relative deviations
below the mean $\langle\bar{h}$$-$$h_{\min}\rangle$, and the maximum
spread $\langle h_{\max}$$-$$h_{\min}\rangle$ follow the same
scaling with the system size $N$. Note, that for our specific system
(PDES time horizon), the ``microscopic'' dynamics is inherently
non-linear, but the effects of the non-linearities only give rise to a
renormalized mass $\Sigma(p)$ (leaving $\Sigma(p)$$>$$0$ for all
$p$$>$$0$) \cite{KORNISS03a,future}. Thus, the dynamics is effectively
governed by relaxation in a small world, yielding a
finite correlation length and, consequently, the slow logarithmic
increase of the extreme fluctuations with the system size
[Eq.~(\ref{mean_max})]. 
Also, for the PDES time horizon, the local
height distribution is asymmetric with respect to the mean, but the
average size of the height fluctuations is, of course, finite for both
above and below the mean. This specific characteristic simply
yields different prefactors for the extreme fluctuations
[Eq.~(\ref{mean_max})] above and below the mean, leaving the
logarithmic scaling with $N$ unchanged. 
%%%%%%%%%%%%%%%%%%%%%%%%%%%%%%%%%%%%%%%%%%%%%%%%%%%%%%%%%%%%%%%%%%%%%
%%%%%%%%%%%%%%%%%%%%%%%%%%%%%%%%%%%%%%%%%%%%%%%%%%%%%%%%%%%%%%%%%%%%%%%%

In summary, we considered the extreme-height fluctuations in a prototypical
model with local relaxation, unbounded local variables, and
short-tailed noise. We argued, that when the interaction topology is
extended to include random links in a SW fashion, the statistics of
the extremes is governed by the FTG distribution. This finding
directly addresses synchronizability in generic SW-coupled systems
where relaxation through the links is the relevant node-to-node
process and effectively governs the dynamics. We illustrated our
results on an actual 
synchronizational problem in the context of scalable parallel simulations.
Analogous questions for heavy-tailed noise distribution and different
types of networks have relevance to various transport and transmission
phenomena in natural and artificial networks \cite{flux}. For example,
heavy-tailed noise typically generates similarly tailed local
field variables through the collective dynamics. Then, the largest
fluctuations can still diverge as a power law with the system size
(governed by the Fr\'echet distribution \cite{GUMBEL,GALAMBOS}),
motivating further research for the properties of
extreme fluctuations in complex networks \cite{future}.

%%%%%%%%%%%%%%%%%%%%%%%%%%%%%%%%%%%%%%%%%%%%%%%%%%%%%%%%%%%%%%%%%%%%%%

We thank Z. R\'acz, Z. Toroczkai, M.A. Novotny, and A. Middleton for
comments and discussions. G.K. thanks CNLS LANL for their hospitality
during Summer 2003. Supported by NSF Grant No. DMR-0113049 and the
Research Corp. Grant No. RI0761. H.G. was also supported in part by
the LANL summer student program in 2003 through US DOE Grant No. W-7405-ENG-36.

%%%%%%%%%%%%%%%%%%%%%%%%%%%%%%%%%%%%%%%%%%%%%%%%%%%%%%%%%%%%%%%%%%%%%%%%%%%%%%


\begin{thebibliography}{99}
\bibitem{note3}
The exponent $\delta$ for the tail of the local relative height
distribution may differ from that of the noise as a result of the
collective (possibly non-linear) dynamics, but the exponential-like
feature does not change.
\bibitem{note2}
When comparing with experimental or simulation data, instead of 
Eq.~(\ref{gumbel}), it is often convenient to use the form of the FTG 
distribution which is scaled to zero mean and unit variance, yielding 
$\tilde{P}_{<}^{\max}(y)=e^{-e^{-(ay+\gamma)}}$,
where $a=\pi/\sqrt{6}$ and $\gamma$ is the Euler constant. The
corresponding FTG density is then
$\tilde{p}^{\max}(y)=ae^{-(ay+\gamma)-e^{-(ay+\gamma)}}$.
\end{thebibliography}
\end{document}